\documentclass[english,11pt,nofootinbib,showpacs,notitlepage]{revtex4-1}
\usepackage[T1]{fontenc}
\usepackage{babel}
\usepackage[utf8]{inputenc}
\usepackage{babel,amsmath,amssymb,mathrsfs,amsfonts,calligra,euscript,textcomp,graphicx,wrapfig,fontenc,pst-blur,tabu,lipsum,%
float,empheq,pst-node,environ,framed,ccaption,capt-of,enumitem,subcaption,enumitem,framed,txfonts,lmodern,cancel,diagbox,tabularx,multirow}
\usepackage[section]{placeins}
\definecolor{darkgreen}{rgb}{0,.7,0}
\usepackage[colorlinks,linkcolor=blue,citecolor=green,pagebackref=false]{hyperref}
\usepackage{float}
\def\[{\left[}
\def\]{\right]}
\def\({\left(}
\def\){\right)}

\def\1{{\bf CI}}
\def\2{{\bf CII}}
\def\3{{\bf CIII}}

\newcommand{\eq}[1]{\begin{equation}#1\end{equation}}

\newcommand{\bgp}[1]{\bigl(#1\bigr)}

\newcommand{\lrsb}[1]{\left[#1\right]}

\newcommand{\nd}{\noindent}

\newenvironment{tightcenter}{%
  \setlength\topsep{0pt}
  \setlength\parskip{0pt}
  \begin{center}
}{%
  \end{center}
}

\begin{document}

\title{Splitting into two isotropic subspaces as a result of cosmological evolution in Einstein-Gauss-Bonnet gravity}

\author{Dmitry Chirkov}
\affiliation{Sternberg Astronomical Institute, Moscow State University, Moscow, Russia}
\affiliation{Bauman Moscow State Technical University, Moscow, Russia}
\author{Alexey Toporensky}
\affiliation{Sternberg Astronomical Institute, Moscow State University, Moscow, Russia}
\affiliation{Faculty of Physics, Higher School of Economics, Moscow, Russia}

\begin{abstract}
We consider numerically dynamics of a flat anisotropic Universe in Einstein-Gauss-Bonnet gravity
with positive $\Lambda$ in dimensionalities 5+1 and 6+1. We identify three possible outcomes of the evolution, one singular
and two nonsingular. First nonsingular outcome is oscillatory. Second is the known exponential
solution. The simplest version of it is the isotropic de Sitter solution. In Gauss-Bonnet cosmology
there exist also anisotropic exponential solutions.
When an exponential solution being an outcome of cosmological evolution
has two different Hubble parameters, the evolution leads from
initially totally anisotropic stage to a warped product of two isotropic subspaces. We show that such type
of evolution is rather typical and possible even in the case when de Sitter solution also exists.
\end{abstract}


\maketitle

\section{Introduction}
Cosmological dynamics in Einstein-Gauss-Bonnet (EGB) gravity have been studied intensively during
last three decades. This modification of gravity is the simplest example (apart from General
Relativity (GR) itself) of Lovelock gravity -- the theory which keeps the order of equations
of motion the same as in GR \cite{Low}. In the standard metric formalism the Lovelock gravity is the only
generalisation of GR with this property  -- other modifications like $f(R)$ theory~\cite{Sotiriou} usually lead
to 4-th order equations of motion instead of the 2-nd order in GR. Bigger order of equation
leads to bigger dimensionality of the corresponding phase spaces of cosmological solutions,
and as a result, many new classes of solutions impossible in GR appear in different versions
of $f(R)$ as well as other 4-order theories. As for Lovelock gravity, it is reasonable to expect
less drastic changes with respect to GR. Indeed, some of rather unusual for GR regimes like
isotropic vacuum solution and its modifications \cite{iso1,iso2,iso3} do not present in Lovelock gravity.

However, rather recently a class of solutions with no direct analog have been found in Einstein-Gauss-Bonnet cosmology \cite{Iv-10} and later have been generalized to Lovelock gravity
\cite{ChPavTop1}. In this class scale
factors change exponentially with time, so Hubble parameters remain constant. In GR there are no
vacuum solutions of this type \cite{PT}, and the only solution with non-zero cosmological constant
$\Lambda>0$ is the de Sitter solution, when all Hubble parameters are equal to the same value.
On the contrary, there are anisotropic vacuum solutions in EGB cosmology, as well as anisotropic
solutions with non-zero cosmological constant \cite{IvErn,IvKob,ChPavTop}. In this sense, a positive cosmological constant
in the whole multidimensional space-time does not lead ultimately to isotropization, as it is in GR.

In the vacuum case a nonsingular cosmological evolution of a flat Universe in EGB gravity is usually
directed from Gauss-Bonnet Kasner regime to the classical GR Kasner regime \cite{PTop} (except for a
pathological 4+1 dimension case \cite{Ronid} which will not be considered here). Since Hubble parameters
decrease by absolute values, in the late time of such evolution the influence of Gauss-Bonnet term
is negligible. On the contrary, if the cosmological dynamics tends to an exponential solution, the
influence of GB term does not disappear at the late time, since Hubble parameters tend to constants. In the present
paper we focus on the case with positive $\Lambda$ in the action, so a late-time Kasner GR solution
does not exist for our set-up, and we expect that there are exponential solutions which dominate
in the future asymptotics of the cosmological evolution.

Despite the fact that there exist anisotropic exponential solutions in the presence of positive $\Lambda$,
it is known that
 all Hubble parameters can not be different. It have been shown that in EGB gravity
it can be no more than 3 different Hubble parameters independently of the number of dimensions \cite{Iv-16}.
This means that a solution in question represent a division of the space into a warped product
of two or three isotropic subspaces. Note, that such a decomposition, in particular case with two
isotropic subspaces, one expanding and one contracting, is often appears as a starting point in papers
devoted to compactification studies. Assuming only two different scale factors allows to get many
results in analytic form despite of big number of dimension. On the contrary, if exponential solutions
can be dynamical attractors for initially totally anisotropic geometries, we get this decomposition
in a natural way dynamically. To study this possibility, first we need get results on stability
of these solutions. The analysis of \cite{Pavl-15,ErIvKob-16,ChT}    indicate that solutions with no one-dimensional subspaces
are locally stable when sum of Hubble-like parameters is positive, except for some special discrete sets of coupling constant. However, local stability
is not  enough to claim that these solutions are natural attractors for general anisotropic initial
conditions. To claim that we need a global analysis. This is the goal of the present paper.

\section{The set-up}
We consider the Einstein-Gauss-Bonnet action in $(D+1)$-dimensional spacetime $\mathcal{M}$:
\eq{S=\int_{\mathcal{M}}d^{D+1}x\sqrt{|g|}\left\{R-\Lambda+\alpha\bgp{R_{\mu\nu\rho\sigma}R^{\mu\nu\rho\sigma}-4R_{\mu\nu}R^{\mu\nu}+R^2}\right\},}
where $|g|$ is the determinant of metric tensor, $R,R_{\alpha\beta},R_{\alpha\beta\gamma\delta}$ are the $(D+1)$-dimensional scalar curvature, Ricci tensor and Riemann tensor respectively; $\alpha$ is the coupling constant; $\Lambda$ is the cosmological term.

We introduce the metric \emph{ansatz} as
\eq{ds^2=-dt^2+\sum\limits_{k=1}^D a^2_k(t)dx_k^2}
In spatially flat model scale factors $a_k(t)$ are defined only up to a constant multiple, so it makes sense to use the Hubble parameters $H_k=\frac{\dot{a}_k}{a_k},\;k=1,\ldots,D$ instead of scale factors.

Equations of motion read
\eq{\begin{split}
      2\sum\limits_{i\ne j}(\dot{H}_i+H_i^2)&+2\sum\limits_{\{i>k\}\ne j}H_i H_k+ \\
        & +8\alpha\sum\limits_{i\ne j}(\dot{H}_i+H_i^2)\sum\limits_{\{k>l\}\ne\{i, j\}}H_k H_l+24\alpha\sum\limits_{\{i>k>l>m\}\ne j}H_i H_k H_l H_m=\Lambda,\quad j=1,\ldots,D
    \end{split}\label{eq.of.motion}}
In addition to the equations of motion we have the constraint equation:
\eq{2\sum\limits_{i>j}H_i H_j+24\alpha\sum\limits_{i>j>k>l}H_i H_j H_k H_l=\Lambda\label{constraint}}
In what following we consider only stable exponential solutions and assume that $\sum\limits_{i=1}^D H_i > 0$.

\section{Numerical analysis of solutions for 5+1 and 6+1 dimensional models}
For numerical calculations we choose sets of initial values of Hubble parameters with different numbers of positive and negative elements. and denote a set with $p$ positive and $n$ negative elements by $(p-n)$; it implies that initially $p$ dimensions are expanding, $n$ dimensions are contracting. We use this notation to simplify comparing the number of expanding dimensions at the beginning and at the end of integration, in case when the outcome of the dynamics studied is an exponential solution. We will see later that the change in number of expanding dimensions is possible indeed.

In all our numerical experiments initial values of Hubble parameters is chosen at random from (-10;10) with uniform probability; totally we have checked up $\sim 10^4$ sets of initial values for each dimensionality. In addition we choose only such initial values of Hubble parameters that sum of them is positive. Since one of the Hubble parameters is defined from the constraint, in some cases we could not find any sets of initial values with a certain number of positive and negative Hubble parameters which obey this condition; for example, for $D=5,\,\alpha>0$ we found no sets of initial values with 1 positive, 4 negative Hubble parameters and initial values with all 5 positive Hubble parameters such that the sum of them is positive. In such cases, the tables demonstrating distribution of results of numerical calculations do not contain the corresponding rows (it is tables~\ref{D=5,alpha=1},\ref{alpha=1},\ref{alpha=0.1}).

We consider 5+1 and 6+1 dimensional models; each of these models has several classes of stable exponential solutions with isotropic subspaces. A solution containing $k$-dimensional and $m$-dimensional isotropic subspaces is referred to as $[k,m]$-splitting in what follows. We use braces when we need to specify whether isotropic subspaces are expanded or contracted; if an isotropic subspace is expanded we use positive number; if an isotropic subspace is contracted we use negative number. For example, notation $\{2,-3\}$ means that the space representing by a corresponding solution is a warped product of 3D contracting isotropic subspace and 2D expanding isotropic subspace. Such way, a case with $[k,m]$-splitting includes 4 subcases: $\{k,m\},\{k,-m\},\{-k,m\},\{-k,-m\}$.

Apart from exponential solution the cosmological dynamics in question can end up in a non-standard singularity,
when time derivative of Hubble parameters diverge while they themselves are finite. In 6+1 dimensions
we have found also one more regime which will be described later in the corresponding section.
\subsection{5+1 dimensional model}
\nd There are 2 classes of stable exponential solutions in EGB model for $D=5$~\cite{ChPavTop}:
\begin{enumerate}
  \item Solution with [3,2]-splitting ($H_1=H_2=H_3=H,\;H_5=H_6=h$) is defined by the equations
  \eq{3\phi^3-7\phi^2+(1+4\alpha\Lambda)\phi-1=0,\;\;\phi=4\alpha H^2;\quad h=-\frac{H}{2}-\frac{1}{8\alpha H}\label{3+2}}
  Explicit solution for the first of these equations is cumbersome enough, so we do not write it down here; in what following we discuss briefly existence conditions for solutions of this equation.
  \item Isotropic solution with $H_1=H_2=H_3=H_4=H_5=H_6=H$ and
  \eq{H^2=-\frac{1}{12\alpha}\lrsb{1\pm\sqrt{1+\frac{6\alpha\Lambda}{5}}}\label{5+0}}
  Solving the latter equation we choose $H>0$ (for stability).
\end{enumerate}
It is easy to check that when $\alpha>0$ there exist both solution with [3,2]-splitting and isotropic solution for any $\Lambda>0$. This is confirmed numerically: table~\ref{D=5,alpha=1} shows what we get starting from initial values with different splittings. We list only solutions existing for the set of coupling studied, zeros in the table do not necessary mean that the corresponding solution can not be reached from
the initial conditions indicated, but it can mean that we have insufficient statistics to detect it.

\begin{table}[!h]
  \caption{The percentage ratio of solutions of different classes for $\alpha=1,\,\Lambda=1;\;D=5$}
  \label{D=5,alpha=1}
  \begin{tabular}{|c|c|c|c|c|c|}
  \hline
  \diagbox[width=10em]{Initial\\splitting}{\\Result} & \{2,-3\} & isotropic & non-standard singularity  \\
  \hline
  (2-3) & 34.6 & 0 & 65.4 \\
  \hline
  (3-2) & 8.6 & 11.5 & 79.9 \\
  \hline
  (4-1) & 5.3 & 69.1 & 25.6 \\
  \hline
  \end{tabular}
  \end{table}

In the case $\alpha<0$ solutions with [3,2]-splitting exist iff $|\alpha|\Lambda>\frac{5}{6}$; isotropic solutions exist under condition $|\alpha|\Lambda<\frac{5}{6}$. Numerical calculations verify these conclusions (see tables~\ref{D=5,alpha=-1,lambda=0.5} and~\ref{D=5,alpha=-1,lambda=1}).

\begin{table}[!h]
  \caption{The percentage ratio of solutions of different classes for $\alpha=-1,\,\Lambda=0.5;\;D=5$}
  \label{D=5,alpha=-1,lambda=0.5}
  \begin{tabular}{|c|c|c|c|c|c|}
  \hline
  \diagbox[width=10em]{Initial\\splitting}{\\Result} & isotropic & non-standard singularity  \\
  \hline
  (1-4) & 0 & 100.0 \\
  \hline
  (2-3) & 0 & 100.0 \\
  \hline
  (3-2) & 0 & 100.0 \\
  \hline
  (4-1) & 37.0 & 63.0 \\
  \hline
  \end{tabular}
  \end{table}

\begin{table}[!h]
  \caption{The percentage ratio of solutions of different classes for $\alpha=-1,\,\Lambda=1;\;D=5$}
  \label{D=5,alpha=-1,lambda=1}
  \begin{tabular}{|c|c|c|c|c|c|}
  \hline
  \diagbox[width=10em]{Initial\\splitting}{\\Result} & \{3,2\} & non-standard singularity  \\
  \hline
  (1-4) & 0 & 100.0 \\
  \hline
  (2-3) & 94.9 & 5.1 \\
  \hline
  (3-2) & 77.4 & 22.6 \\
  \hline
  (4-1) & 44.0 & 56.0 \\
  \hline
  \end{tabular}
  \end{table}

\subsection{6+1 dimensional model}
\nd Below we list 3 classes of stable exponential solutions in EGB model for $D=6$~\cite{ChPavTop1}:
\begin{enumerate}
  \item Solution with [3,3]-splitting~\cite{IvKob-2}, where $H_1=H_2=H_3=H,\;H_4=H_5=H_6=h$ and
  \eq{\Lambda=\frac{6(H^4+4H^3h+5H^2h^2+4Hh^3+h^4)}{H^2+4Hh+h^2},\quad h=-2H\pm\sqrt{3H^2-\frac{1}{4\alpha}}\label{3+3}}
  \item Solution with [4,2]-splitting~\cite{IvKob-3}, where $H_1=H_2=H_3=H_4=H,\;H_5=H_6=h$ and
  \eq{\Lambda=\frac{2(5H^3+6H^2h+3Hh^2+h^3)}{H+h},\quad h=-H-\frac{1}{12\alpha H}}
  \item Isotropic solution with $H_1=H_2=H_3=H_4=H_5=H_6=H$ and
  \eq{H^2=-\frac{1}{24\alpha}\lrsb{1\pm\sqrt{1+\frac{8\alpha\Lambda}{5}}}}
  Solving the latter equation we choose $H>0$ (for stability).
\end{enumerate}
Below we discuss solutions with $\alpha>0$ and $\alpha<0$ separately.

\subsubsection{The case $\alpha>0$}
First, we find out for what values of $\alpha>0$ and $\Lambda>0$ there exist solutions with [3,3]- and [4,2]-splittings as well as isotropic solution, and are there such $\alpha$ and $\Lambda$ that all these solutions  co-exist, i.e. we can reach each of these solutions for different initial values of Hubble parameters but for the same values of $\alpha$ and $\Lambda$.

\nd\textbf{I. [3,3]-splitting}: substituting the expression for $h$ into the equation for $\Lambda$ and doing simple transformations we obtain
\eq{20736\,{z}^{4}-17280\,{z}^{3}+ \left( 2736+4032\,\xi \right) {z}^{2}-\left(216+528\,\xi \right) z+ \left(2\,\xi+3 \right)^{2}=0}
where $z=H^2\alpha,\;\xi=\Lambda\alpha$. Note that $z>0$ and $\xi>0$. Solutions of this equation are
\eq{z={\frac {5}{24}}\pm \frac{\sqrt {1-2\,\xi}}{6}\pm \frac{1}{24}\sqrt {-{\frac {24\,\xi\,\sqrt {1-2\,\xi}-21\,\sqrt {1-2\,\xi}+48\,\xi-24}{\sqrt {1-2\,\xi}}}}}
There exists at least one positive $z$ for all $\xi<\frac{1}{2}$.

\nd\textbf{II. [4,2]-splitting}: substituting the expression for $h$ into the equation for $\Lambda$ we get
\eq{f_1(H)=0,\;\;\mbox{where}\;\;f_1(H)=1728\,{H}^{6}\alpha^{6}-432\,{H}^{4}\alpha^{4}+72\,\alpha^{4}{H}^{2}\Lambda-1}
This equation has at least one positive root for any values $\alpha$ and $\Lambda$ since $f_1(0)=-1$ and $f_1(H)\stackrel{H\rightarrow+\infty}{\longrightarrow}+\infty$.

\nd\textbf{III. isotropic solution} exists for any $\alpha>0$ and $\Lambda>0$:
\eq{H=\sqrt{-\frac{1}{24\alpha}\lrsb{1-\sqrt{1+\frac{8\alpha\Lambda}{5}}}}}
  \begin{table}[!h]
  \caption{The percentage ratio of solutions of different classes for $\alpha=1,\,\Lambda=10;\;D=6$}
  \label{alpha=1}
  \begin{tabular}{|c|c|c|c|c|c|}
  \hline
  \multirow{2}{*}{\diagbox[height=2\line]{Initial splitting}{Result}} & \multirow{2}{*}{\{2,-4\}} & \multirow{2}{*}{isotropic} & \multicolumn{2}{c|}{oscillations} & \multirow{2}{*}{$\begin{array}{c}
                                                        \mbox{non-standard} \\
                                                        \mbox{singularity}
                                                      \end{array}$} \\
  \cline{4-5}
   &  &  & \{1,1,-1,-1,-1,-1\} & \{1,1,1,1,-1,-1\} & \\
  \hline
  (2-4) & 0.1 & 0 & 5.5 & 0 & 94.4 \\
  \hline
  (3-3) & 1.3 & 0 & 0 & 0 & 98.7 \\
  \hline
  (4-2) & 0.4 & 11.0 & 0 & 0.2 & 88.4 \\
  \hline
  (5-1) & 0.5 & 94.5 & 0 & 0 & 5.0 \\
  \hline
  (6-0) & 2.7 & 67.6 & 0 & 0 & 29.7 \\
  \hline
  \end{tabular}
  \end{table}

  \begin{table}[!h]
  \caption{The percentage ratio of solutions of different classes for $\alpha=0.1,\,\Lambda=2.5;\;D=6$}
  \label{alpha=0.1}
  \begin{tabular}{|c|c|c|c|c|}
  \hline
  \diagbox[width=10em]{Initial\\splitting}{\\Result} & \{3,-3\} & \{4,-2\} & isotropic & non-standard singularity \\
  \hline
  (2-4) & 0 & 0 & 0 & 100.0 \\
  \hline
  (3-3) & 19.1 & 23.6 & 43.3 & 14.0 \\
  \hline
  (4-2) & 0 & 35.3 & 45.8 & 18.9 \\
  \hline
  (5-1) & 0 & 0 & 92.5 & 7.5 \\
  \hline
  (6-0) & 0 & 0 & 97.6 & 2.4 \\
  \hline
  \end{tabular}
  \end{table}
Summing things up we can say that solutions of all three types co-exist for $0<\alpha\Lambda<\frac{1}{2}$; for $\alpha\Lambda\geqslant\frac{1}{2}$ only isotropic solution and solution with [4,2]-splitting exist. This conclusions are confirmed by the results of numerical calculations for $\alpha=1,\,\Lambda=10$ (see table~\ref{alpha=1}) and $\alpha=0.1,\,\Lambda=2.5$ (see table~\ref{alpha=0.1}).

\begin{figure}[!h]
\begin{minipage}[h]{.49\linewidth}
\center{\includegraphics[width=\linewidth]{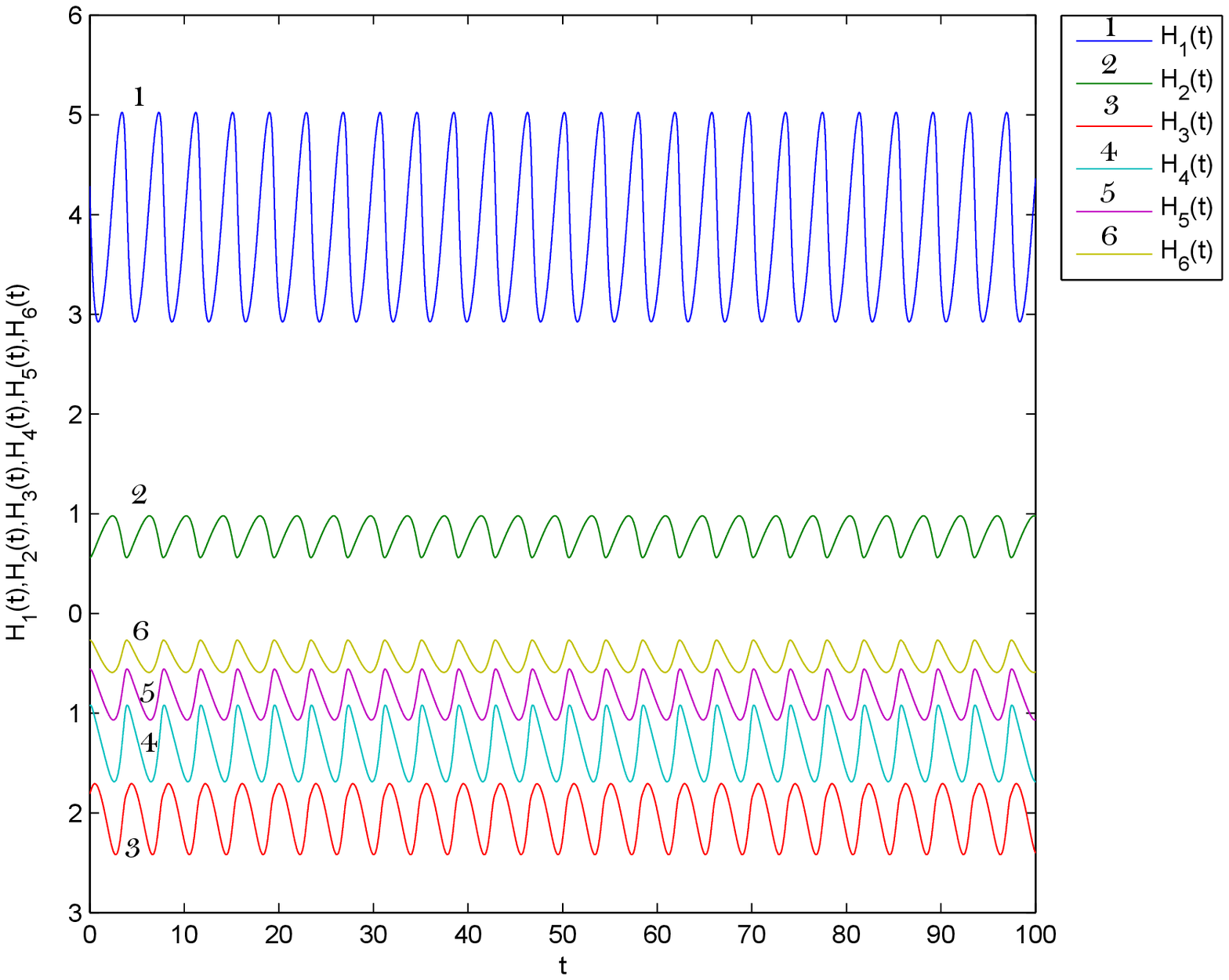} \\ a)}
\end{minipage}
\hfill
\begin{minipage}[h]{.49\linewidth}
\center{\includegraphics[width=\linewidth]{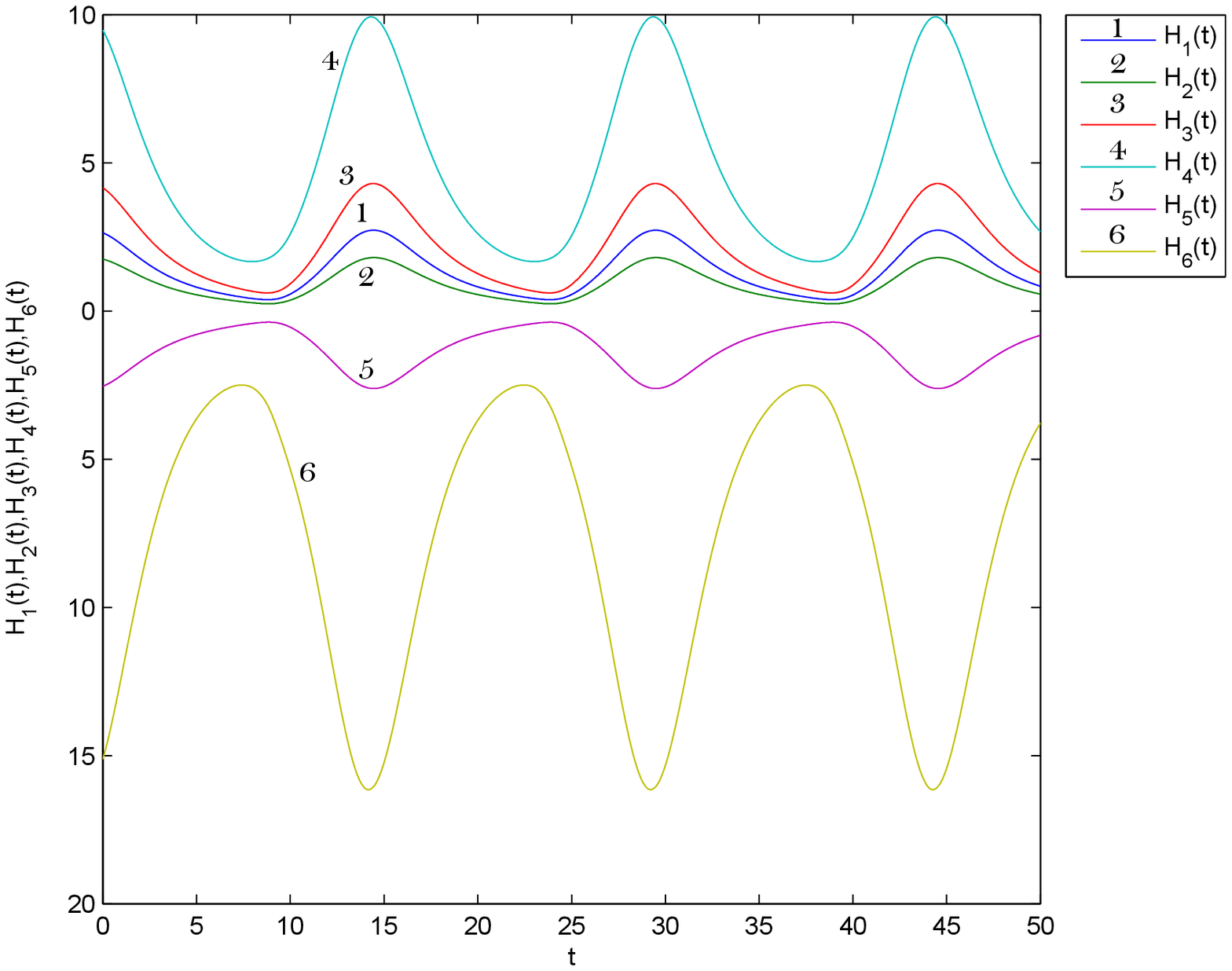} \\ b)}
\end{minipage}
\caption{\footnotesize Numerical solutions for $D=6,\;\alpha=1,\;\Lambda=10$. The figure a) depicts oscillatory solution with \{1,1,-1,-1,-1,-1\}-splitting; the figure b) demonstrates oscillatory solution with \{1,1,1,1,-1,-1\}-splitting.}
\label{oscillations}
\end{figure}

Numerical calculations show the existence of some specific solutions which we call "oscillations". We found oscillations in 6+1 dimensional model with $\alpha>0$ and for large enough values of $\alpha\Lambda$ (see Fig.~\ref{oscillations}). Numerical calculations show also that despite the overall measure of initial
conditions set leading to oscillations is not large, the oscillation regime is stable with respect to
small perturbations of initial conditions. The period of oscillations is fixed for fixed coupling constants,
while amplitudes can vary depending on initial conditions.
 Note that there are no oscillations  found for $D=5$ case.

\subsubsection{The case $\alpha<0$}
Analogously to the previous subsection we find out what solutions can exist for negative values of $\alpha$ (as before, we assume that $\Lambda>0$).

\nd\textbf{I. [3,3]-splitting}: substituting the expression for $h$ into the equation for $\Lambda$ we obtain
\eq{20736\,{z}^{4}+17280\,{z}^{3}+ \left( 2736-4032\,\xi \right) {z}^{2}+\left( 216-528\,\xi \right) z+ \left( 2\,\xi-3 \right) ^{2}=0}
where $z=H^2\alpha,\;\xi=\Lambda|\alpha|$. Note that $z>0$ and $\xi>0$ again. Solutions of this equation are
\eq{z=-{\frac {5}{24}}\pm \frac{\sqrt {1+2\,\xi}}{6}\pm \frac{1}{24}\sqrt {{\frac {24\,\xi\,\sqrt {1+2\,\xi}+21\,\sqrt {1+2\,\xi}-48\,\xi-24}{\sqrt {1+2\,\xi}}}}}
It is easy to check that we have at least one positive $z$ for all $\xi\geqslant\frac{5}{8}$.

\nd\textbf{II. [4,2]-splitting}: substituting the expression for $h$ into the equation for $\Lambda$ we get
\eq{f_2(z)=0\;\;\mbox{where}\;\; f_2(z)=1728\,{z}^{3}|\alpha|^3-72\,\Lambda\,\alpha^2z+432\,\alpha^2{z}^{2}+1,\;\;z\equiv H^2}
One can see that $f_2(0)=1,\;f_2(z)\stackrel{z\rightarrow+\infty}{\longrightarrow}+\infty$; therefore, in order to the equation $f_2(z)=0$ has at least one positive root it is necessary that the function $f_2$ has non-positive minimum at a point $z_+>0$. It is easy to check that minimum point of the function $f_2$ is
\eq{z_+=\frac {-1+\sqrt {2\,\Lambda\,|\alpha|+1}}{12|\alpha|}>0\;\;\mbox{for any values of}\;\alpha,\,\Lambda}
Substituting that $z_+$ into $f_2$ we obtain
\eq{f_2^{\rm min}=-4\,\xi\,\sqrt {2\,\xi+1}+6\,\xi-2\,\sqrt {2\,\xi+1}+3,\quad\xi=\Lambda|\alpha|}
It is easy to prove that $f_2^{\rm min}\leqslant0$ for all $\xi\geqslant\frac{5}{8}$.

\nd\textbf{III. isotropic solution} exists for such values of $\alpha$ and $\Lambda$ that $|\alpha|\Lambda\equiv\xi\leqslant\frac{5}{8}$
\eq{H=\sqrt{\frac{1}{24|\alpha|}\lrsb{1\pm\sqrt{1-\frac{8|\alpha|\Lambda}{5}}}}}
So, we see that for $|\alpha|\Lambda>\frac{5}{8}$ only [3,3]-splitting and [4,2]-splitting exist (see table~\ref{alpha=-1}); for $|\alpha|\Lambda<\frac{5}{8}$ only isotropic solution exists (see table~\ref{alpha=-0.3}).
  \begin{table}[!h]
  \caption{The percentage ratio of solutions of different classes for $\alpha=-1,\,\Lambda=10;\;D=6$}
  \label{alpha=-1}
  \begin{tabular}{|c|c|c|c|c|}
  \hline
  \diagbox[width=10em]{Initial\\splitting}{\\Result} & \{4,-2\} & \{4,2\} & \{3,-3\} & non-standard singularity \\
  \hline
  (1-5) & 0 & 0 & 0 & 100.0 \\
  \hline
  (2-4) & 0 & 0.1 & 0 & 99.9 \\
  \hline
  (3-3) & 24.9 & 3.0 & 19.7 & 52.4 \\
  \hline
  (4-2) & 67.4 & 0.4 & 0.2 & 32.0 \\
  \hline
  (5-1) & 1.2 & 0 & 0 & 98.8 \\
  \hline
  (6-0) & 0 & 96.9 & 0.8 & 2.3 \\
  \hline
  \end{tabular}
  \end{table}

\begin{table}[!h]
  \caption{The percentage ratio of solutions of different classes for $\alpha=-0.3,\,\Lambda=1;\;D=6$}
  \label{alpha=-0.3}
  \begin{tabular}{|c|c|c|c|c|}
  \hline
  \diagbox[width=10em]{Initial\\splitting}{\\Result} & isotropic & non-standard singularity \\
  \hline
  (1-5) & 0 & 100.0 \\
  \hline
  (2-4) & 0 & 100.0 \\
  \hline
  (3-3) & 3.0 & 97.0 \\
  \hline
  (4-2) & 1.5 & 98.5 \\
  \hline
  (5-1) & 39.0 & 61.0 \\
  \hline
  (6-0) & 74.0 & 26.0 \\
  \hline
  \end{tabular}
  \end{table}

\section{Conclusions}
We have considered a cosmological evolution of a flat 5+1 and 6+1 dimensional anisotropic Universe in Gauss-Bonnet gravity.
We started from an arbitrary anisotropic initial conditions and study the outcome of corresponding
cosmological evolution. Three possible outcomes have been identified. About a half of trajectories traced (we have $\sim 10^4$ trajectories for each dimensionality) ends in a non-standard singularity. A few percents of trajectories represent a periodic
in Hubble parameters solution which, to our knowledge, have not been described before. And the rest of
them (roughly a half of overall number) lead to exponential solutions.

For the studied dimensionality only exponential solutions with one or two different Hubble parameters are
stable. In the former case we have a de Sitter solution. In the latter case
 initially totally anisotropic Universe becomes a warped product of two isotropic
subspaces. What follow from our results is that this can happen even if de Sitter solution exists for
the particular set of coupling constants -- nevertheless, depending on initial conditions, a Universe
can "miss" de Sitter stage and split into two isotropic subspaces.
We can see also that despite most trajectories have conserved number of expanding and contracting
dimensions, nevertheless
 signs of Hubble parameters can change during cosmological evolution.

Our results show that a situation when a space is splitted into a warped product of isotropic subspaces
can be a natural result in cosmological evolution of a flat Universe in Gauss-Bonnet gravity.

 {\bf Acknowledgments}

The work of A.T. is supported by RFBR grant 17-02-01008. Authors are grateful to Sergey Pavluchenko
for discussions.

\end{document}